\documentclass[12pt]{iopart}

\usepackage{iopams}
\usepackage{graphicx}
\usepackage{color}
\usepackage{bm}

\newcommand{\beq}{\begin{equation}}
\newcommand{\eeq}{\end{equation}}
\newcommand{\beqa}{\begin{eqnarray}}
\newcommand{\eeqa}{\end{eqnarray}}

\usepackage{dcolumn}
\usepackage{epsf}
\usepackage{verbatim}

\newcommand{\edc}{\end{document}}
\newcommand{\bb} {\begin{thebibliography}}
\newcommand{\eb}{\end{thebibliography}}
\newcommand{\bc}{\begin{center}}
\newcommand{\ec}{\end{center}}

\newcommand{\be}{\begin{equation}}
\newcommand{\ee}{\end{equation}\normalsize}
\newcommand{\bea}{\begin{eqnarray}}
\newcommand{\eea}{\end{eqnarray}}
\newcommand{\ba}{\begin{array}{l}}
\newcommand{\lab}[1]{\label{#1}}
\newcommand{\ea}{\end{array}}
\newcommand{\dsfrac}{\displaystyle\frac}
\newcommand{\ds} {\displaystyle}

\newcommand{\re}[1]{(\ref{#1})}

\newcommand{\ci}{\cite}

\newcommand{\dsint}{\ds\int}

\newcommand{{\vergul}}{  ,}

\newcommand{\veps}{\varepsilon }

\newcommand{\req}[1]{(\ref{#1})}
\def\cE{{\cal E}}

\begin{document}
\title{The effects of disorder in dimerized quantum magnets in mean field approximations}
\author{Abdulla Rakhimov$^{a,b}$, Shuhrat Mardonov$^{c}$, E. Ya. Sherman$^{d,e}$, Andreas Schilling$^{b}$}
\address{
$^a$ Institute of Nuclear Physics, Tashkent 100214, Uzbekistan\\
$^b$ Physik-Institut, University of Z\"{u}rich, Winterthurerstrasse 190, 8057 Z\"{u}rich, Switzerland\\
$^c$ Samarkand State University, 703004 Samarkand, Uzbekistan\\
$^d$ Department of Physical Chemistry, University of Basque Country UPV/EHU, 48080 Bilbao, Spain\\
$^e$ IKERBASQUE Basque Foundation for Science, 48011 Bibao, Bizkaia, Spain
}

\ead{evgeny$_{-}$sherman@ehu.es}

\newpage

\begin{abstract}
We study theoretically the effects of disorder on Bose-Einstein condensates (BEC) of bosonic triplon quasiparticles
in doped dimerized quantum magnets.  The condensation occurs in a strong enough magnetic field,
where the concentration of bosons in the random potential is sufficient to form the condensate.
The effect of doping is partly modeled by a $\delta$ - correlated distribution of impurities,
which (i) leads to a uniform renormalization of the system parameters and (ii) produces
disorder in the system with renormalized parameters.
This approach can explain qualitatively the available magnetization data
on the Tl$_{1-x}$K$_{x}$CuCl$_{3}$ compound taken as an example.
In addition to the magnetization, we found that the speed of the Bogoliubov mode has a maximum as a function
of  $x$. No evidence of the pure Bose glass phase has been found in the BEC regime.
\end{abstract}

\newpage

\section{Introduction}

The effects of disorder on the properties of Bose-Einstein condensates present interesting problems
both for theoretical and experimental physics \cite{pelster,vinokur,giorgini,gaul,zhang,graham}.
Disorder is important in various systems of real particles
such as superfluid $^4$He, cold atoms in optical lattices, and quasiparticles such as
polaritons \cite{shelykh} and excitons \cite{butov}.
These systems are well-suited for experimental studies, however the theory
of disordered ensembles of interacting bosons is complex and there are essentially
no exact solutions even in one dimension \ci{shapiro}.
To approach this problem, Yukalov and Graham (YG) developed a self-consisted stochastic
mean field approximation (MFA) \ci{yukgraham} for Bose systems with arbitrarily strong interparticle
repulsion and arbitrary strength of disorder potential. It was shown that, in general, the Bose system consists
of following coexisting components: the condensate fraction, $\rho_0$, the normal fraction $\rho_N$,
the glassy fraction $\rho_G$,
and, in addition, can be characterized by the superfluid density $\rho_s$.
In the limit of asymptotically weak interactions and disorder
the known results, obtained in pioneering work by Huang and Meng \ci{huangmeng} (HM) are reproduced
by the YG theory. An interesting question here concerns the problem about the
existence of a pure Bose glass (BG) phase, i.e.  the phase where the condensate fraction is nonzero, while
the superfluid fraction is not yet present. Note that Ref.\cite{graham} introduced an alternative definition of the gapless BG phase,  
having localized short-lived excitations and vanishing superfluid density 
with a continuous transition to the  normal phase at finite temperature.
Even without disorder, the condensate is depleted by particle-particle interactions and temperature. 
The inclusion of random fields depletes the
condensate further and, possibly, creates the glassy fraction.

As it was understood recently, a new class of BEC can be provided by spin-related quasiparticles
in magnetic solids such as intensively pumped magnons \cite{demokritov} or triplons in the dimerized
quantum magnets in the equilibrium \cite{giamarchi}.
In the magnets, the effect of disorder, which can be produced by admixing other chemical
elements, can be rather strong to be seen in the physical properties 
such as  the temperature-dependent magnetization. The so far most investigated
compound showing BEC of triplons is TlCuCl$_{3}$. To study the effect of disorder,
solid solutions of quantum antiferromagnets  TlCuCl$_{3}$  and  KCuCl$_{3}$, i.e.
Tl$_{1-x}$K$_{x}$CuCl$_{3}$ have been experimentally investigated recently \ci{oosawa65,yamadaglass,tanakaptp}
at low temperatures $T$. The zero-field ground states of  TlCuCl$_{3}$  and KCuCl$_{3}$
are spin singlets with excitation gaps $\Delta_{\rm st}=7.1$ K and $\Delta_{\rm st}=31.2$ K, respectively and the
magnetic excitations are spin triplets.  Triplons arise in magnetic fields $H>H_{c}$, where $H_{c}$
is defined by condition of closing the gap by the Zeeman splitting, that is $\Delta_{\rm st}=g\mu_B H_c$,
where $g$ is the electron Land\'{e} factor and $\mu_B$ is the Bohr magneton.
In the mixture Tl$_{1-x}$K$_{x}$CuCl$_{3}$ the induced magnetization $M$ exhibits  a cusplike minimum at a critical
temperature $T_c(H)$ for fixed magnetic field $H\geq H_{c}$ similarly to the parent compound, and
can be successfully explained in terms of triplon BEC \ci{Yamada08,ourmagnon,Amore08}.

For a theoretical description it is natural to assume that for weak doping
$x\ll1$ in the mixed system  Tl$_{1-x}$K$_{x}$CuCl$_{3}$ a
small admixture of potassium forms a disorder potential. Consequently, the recently developed theories
of "dirty bosons" \ci{pelster,vinokur,yukgraham,huangmeng} can be applied to study the BEC of triplons in Tl$_{1-x}$K$_{x}$CuCl$_{3}$.
Here the following natural questions arise. For example, what is the correspondence
between admixing parameter $x$ and the properties of the disorder potential ? What are the experimental consequences
of the disorder ? Yamada {\it et al.} \ci{yamadaglass}  analyzed the electron spin resonance
spectrum in Tl$_{1-x}$K$_{x}$CuCl$_{3}$ and concluded that there is a Bose glass - BEC transition near a
critical magnetic field. Although this interpretation might need a further analysis (see discussion in Refs. \ci{discusplr})
it would be interesting to study the influence of the glassy phase, or more exactly, of
the glassy fraction $\rho_G$ on the magnetization. Note that even  the existence of a pure Bose glass phase
still is a matter of debate even in  theoretical approaches.
For example, it may be predicted by the approach used by Huang and Meng
\ci{huangmeng} if one extends their formulas from weak disorder to a strong one.  On the other hand,
no pure Bose glass was found in Monte - Carlo simulations \ci{monte} for atomic gases, but predicted for triplons
 at $T=0$ by Nohadani {\it et al.} \ci{nohadani}. 

Here we develop a theory of the disorder effects on the BEC of triplons
taking Tl$_{1-x}$K$_{x}$CuCl$_{3}$ as a prototype for studies of specific properties.
For example, in atomic gases considered in Refs. \ci{yukgraham,huangmeng},
the chemical potential $\mu$ is determined self-consistently with fixed number of atoms, while
in the triplon gas the chemical potential is a given external parameter
controlled by the applied magnetic field and the number
of triplons is conserved in the thermodynamic limit.
To clarify the terms,  we underline that the number of magnons may vary
but that of triplons may be tuned and kept fixed, which makes possible the BEC of the latter.

The paper is organized as follows. In Sections II and III we outline the YG and HM approaches valid only for
$T\leq T_c$  and extend it for the triplon system. The shift of $T_c$
due to disorder and the normal phase properties
will be discussed in Section  IV. Our numerical results will be presented in Section V.
Conclusions will summarize the results of this work.

 \section{Yukalov-Graham approximation for disordered triplons}

In the following we reformulate the Yukalov-Graham approximation to the triplon system with arbitrary disorder.
The Hamiltonian operator of triplons with contact interaction and implemented disorder
potential $V(\mathbf{r})$ is given by
\be
H=\int d^{3}r
\left[
\psi^{\dagger}(\mathbf{r})\left(\hat{K}-\mu+V(\mathbf{r})\right)\psi(\mathbf{r})+
\frac{U}{2}\left(\psi^{\dagger}(\mathbf{r})\psi(\mathbf{r})\right)^{2}
\right],
\label{2.1}
\ee
where $\psi(\mathbf{r})$ is the bosonic field operator, $U$ is the
interparticle interaction strength, and $\hat{K}$ is the kinetic energy operator which
defines the bare  triplon dispersion $\varepsilon_{\mathbf k}$.
Since the triplon BEC occurs in solids, we perform integration over the unit cell of the crystal with  the corresponding
momenta defined in the first Brillouin zone. Below we assume that the bare spectrum
remains coherent in the presence of disorder and consider it  as a simple 
isotropic one: $\varepsilon_{\mathbf k}={k}^2/2m$,
where $m$ is the triplon effective mass. The distribution of random
fields is assumed to be zero - centered, $\left<V(\mathbf{r})\right>=0$, and the correlation function
$R(\mathbf{r}-\mathbf{r}')=\left<V(\mathbf{r})V(\mathbf{r}')\right>$. Here and below we adopt the units $k_B\equiv1$, $\hbar\equiv1$,
and $V\equiv1$ if not stated otherwise for the unit cell volume.

To describe Bose condensed system where the global gauge symmetry is broken, one employs
the Bogoliubov shift:
\be
\psi(\mathbf{r})=\sqrt{\rho_0(\mathbf{r})}+{\psi}_{1}(\mathbf{r})
\lab{2.2}
\ee
where the condensate density $\rho_0(\mathbf{r})$ is constant for the homogeneous system, $\rho_0(\mathbf{r})\equiv\rho_0$.
Since by the definition  the average of $\psi^{\dagger}(\mathbf{r})\psi(\mathbf{r})$ is
the total number of particles:
\be
N=\int_{V}d^{3}r \langle\psi^{\dagger}(\mathbf{r})\psi(\mathbf{r})\rangle
\lab{2.3}
\ee
with the density of triplons per unit cell $\rho=N/V$, from the normalization  condition
\be
\rho=\rho_{0}+\rho_{1}
\lab{2.4}
\ee
one immediately obtains
\be
\rho_{1}=\dsfrac{1}{V}\int_{V}d^{3}r \langle{\psi}_{1}^{\dagger}(\mathbf{r}){\psi}_{1}(\mathbf{r})\rangle.
\lab{2.5}
\ee
Therefore  the field operator ${\psi}_{1}(\mathbf{r})$ determines the density of uncondensed
particles.

The YG approximation is formulated in representative ensemble formalism, which
includes  two Lagrange multipliers, $\mu_0$ and $\mu_1$, defined as:
\be
N_0=-\dsfrac{\partial \Omega}{\partial \mu_0}, \quad
N_1=-\dsfrac{\partial \Omega}{\partial \mu_1},
\label{N0N1}
\ee
where $\Omega$ is the grand thermodynamic potential. It was shown that disorder would not change the explicit
expressions for chemical potentials, obtained earlier \ci{yukannals} in Hartree-Fock-Bogoliubov (HFB) approximation without disorder,
\be
\mu_0=U(\rho+\rho_1+\sigma),\quad
\mu_1=U(\rho+\rho_1-\sigma),
\lab{2.6}
\ee
where $\sigma=\dsfrac{1}{V}\int_{V}d^{3}r \langle{\psi}_{1}(\mathbf{r}){\psi}_{1}(\mathbf{r})\rangle$ is the anomalous density.
The total system chemical potential $mu$, related to the total number of particles as $N=-\partial \Omega/ \partial \mu$, is
determined by
\be
\mu \rho=\mu_1\rho_1+\mu_0 \rho_0.
\lab{2.7}
\ee
Clearly, when the gauge symmetry is not broken, i.e. $\rho_0=0$, $\sigma=0$, $\rho_1=\rho$, both $\mu_0$ and $\mu_1$
coincide giving $\mu=\mu_1=2U\rho$.

In contrast to homogeneous atomic gases considered in Refs.\ci{yukgraham,huangmeng},
where $\rho$ is fixed and $\mu(\rho)$ should be calculated as an output parameter,
in the triplon gas the chemical potential is fixed by the external magnetic field,
while the density $\rho=\rho(\mu)$ should be calculated self consistently.
In fact, in a system of triplons $\mu$ characterizes an additional direct contribution to the triplon energy
due to the external field $H$ and can be written as
\be
\mu=g\mu_B H-\Delta_{\rm st},
\lab{2.8}
\ee
which can be interpreted as a chemical potential of the $S_z=-1$ triplons.

The magnetization is proportional to the triplon density
\be
M=g\mu_B\rho
\lab{2.9}
\ee
with $\rho$ is defined from  \re{2.7} as
\be
\rho=\dsfrac{1}{\mu}\left(\mu_1\rho_1+\mu_0 \rho_0\right)
\lab{2.10}
\ee
where $\mu_0$ and $\mu_1$ are given in \re{2.6} and the densities $\rho_0$, $\rho_1$ must be
calculated self consistently.

It is well known \ci{fisher} that the disorder field leads to formation of a glassy fraction
with the density $\rho_G$. In this approximation each of $\rho_1$ and $\sigma$
are presented as
\be
\rho_1=\rho_N+\rho_G; \quad \sigma=\sigma_N+\rho_G
\lab{2.11}
\ee
where $\rho_N$ and $\sigma_N$ are the normal and anomalous densities without disorder.
In the YG method, based on HFB approximation,
the following explicit relations can be obtained \ci{ourmagnon}:

\begin{eqnarray}
\rho_{N} &=&\dsfrac{(\Delta m)^{3/2}}{3\pi^2}+\dsint \dsfrac{d^3{k}}{\left( 2\pi \right) ^{3}}
f_{B}(\cE_{{k}})
\dsfrac{\veps_{k}+\Delta }{\cE_{k}},
\lab{2.12}\\
\sigma_{N} &=&\dsfrac{(\Delta m)^{3/2}}{\pi^2}-\Delta \dsint \dsfrac{d^3{k}}{\left( 2\pi \right) ^{3}}
f_{B}(\cE_{{k}})\dsfrac{1}{\cE_{{k}}},
\lab{2.13}
\end{eqnarray}
with the Bose distribution of Bogoliubov excitations $f_{B}(\cE_{k})={1}/({e^{\cE_{k}/T}-1})$ having the
dispersion ${\cE_{k}}$
\be
{\cE_{k}}=\sqrt{\veps_k}\sqrt{\veps_k+2\Delta}.
\lab{2.14}
\ee
For small momentum $k$ the dispersion is linear, ${\cE_{k}}=ck$, and the speed of the Bogoliubov mode
\be
c=\dsfrac{\sqrt{\Delta}}{\sqrt{m}}.
\lab{2.15}
\ee

The self energy $\Delta$ is determined formally by the same equation as in the case when the disorder
is neglected,
\be
\Delta=U(\rho_0+\sigma)=U(\rho-\rho_N+\sigma_N).
\lab{2.16}
\ee
The contribution from the disorder potential is hidden in the density of the glassy fraction
\be
\rho_{G}=\dsfrac{1}{V}\int_{V}d^{3}r \langle\langle {\psi}_{1}(\mathbf{r}){\psi}_{1}(\mathbf{r})\rangle\rangle
\lab{2.17}
\ee
where the double angle brackets mean the stochastic average. In general, the calculation of $\rho_G$
is rather complicated, however, for the $\delta$ - correlated disorder i.e. for the white noise,
\be
\langle\langle V(\mathbf{r})V(\mathbf{r}')\rangle\rangle =R\delta(\mathbf{r}-\mathbf{r}'),
\lab{2.18}
\ee
equation \re{2.17} is simplified as \ci{yukgraham}
  \be
  \rho_G=\dsfrac{R_0(\rho-\rho_N)}{R_0+7(1-R_0)^{3/7}}.
  \lab{2.19}
  \ee
The density of condensed fraction can be found by inserting (12) and (20) into the normalization condition \re{2.4}.
The result is
\be
  \rho_0=\dsfrac{7(1-R_0)^{3/7}(\rho-\rho_N)}{R_0+7(1-R_0)^{3/7}}.
  \lab{2.20}
  \ee
In Eqs. \re{2.19} and \re{2.20}
we introduced the dimensionless parameter $R_0$ as
\be
R_0\equiv \dsfrac{7Rm^2}{4\pi\sqrt{m\Delta}}.
\lab{2.21}
\ee
 One can see from Eqs. \re{2.19} and \re{2.20} that the glassy fraction is proportional to the condensed one,
 \be
 \rho_G= \dsfrac{\rho_0 R_0}{7(1-R_0)^{3/7}}.
 \lab{2.22}
 \ee
The system of Eqs. \re{2.6}, \re{2.7}, \re{2.12}-\re{2.18} are the basic of
YG approximation.

 An interesting quantity, crucial for determining the Bose glass phase, is the superfluid density,
 $\rho_s$. In general it is defined as a partial density appearing as a response to a velocity
 boost
 \be
 \rho_s=\dsfrac{1}{3mV}\lim_{\mathbf{v}\rightarrow 0}\dsfrac{\partial}{\partial \mathbf{v}}\langle\hat{\mathbf P}_{\mathbf{v}}\rangle
 \ee
 where $\hat{\mathbf P}_{\mathbf{v}}$ is the total momentum of the system, dependent on the macroscopic velocity ${\mathbf{v}}$.
 Referring  the reader to original papers
 \ci{yukgraham,huangmeng} we bring below analytical expression obtained there for $\rho_s$ in the case
 of white noise random potential
\begin{eqnarray}
\rho_s&=&\rho-\dsfrac{4\rho_G}{3}-\dsfrac{2Q_N}{3T}, \lab{3.7a}\\
Q_N&=&\dsfrac{1}{8m}\dsint\dsfrac{k^2 d^{3}{k}}{(2\pi)^3 \sinh^2 ({\cE_{k}}/2T)}.
\lab{3.7b}
\end{eqnarray}

Note that YG approach is valid for arbitrary strength of
the interaction potential $U$, and for arbitrary strong disorder. For the weak interactions 
it leads to pioneering Huang-Meng approach  \ci{huangmeng},
which will be extended to the ``dirty triplons'' in the next section.

\section{Huang-Meng approximation}
For completeness, we present here the results for the Huang-Meng approach,
based on the so called Hartree Fock Popov (HFP) approximation which has been widely applied in the literature
 to describe the BEC of triplons \ci{Yamada08,Amore08}. The basic equations of this approach
 can be obtained by neglecting the anomalous density $\sigma$, which leads naturally
 to the single chemical potential $\mu=\mu_0=\mu_1$. Namely, one finds from \re{2.6}, \re{2.7}
 and \re{2.16}
\be
\Delta=U\rho_0,\qquad  \mu=U(\rho+\rho_1).
\lab{3.1}
\ee
From these equations and \re{2.11} one obtains following main equations for the self energy $\Delta$:
\begin{eqnarray}
\Delta&=&\mu-2U(\rho_N+\rho_G),
\lab{3.2}
\end{eqnarray}
where $\rho_N$ is formally given in \re{2.12}, and $\rho_0$ is determined by the first equation in (\ref{3.1}).
The glassy fraction can be obtained from \re{2.19} in the linear approximation by $R$
assuming weakness  of interparticle interaction \ci{yukgraham,huangmeng}
\be
\rho_G=\dsfrac{m^2 R}{8\pi^{3/2}}\left(\dsfrac{\rho_0}{a_s}\right)^{1/2},
\lab{3.3}
\ee
where $a_s=Um/4\pi$ is the $s$ - wave scattering length. Inserting \req{3.3} into \req{3.2}   we can rewrite the former as
\be
\Delta=\mu-2U\rho_N-\dsfrac{m^2 R \sqrt{\Delta}}{2\pi \sqrt{m}}.
\lab{3.4}
\ee
To evaluate the densities  one has to solve nonlinear algebraic equation \req{3.4}, where
$\rho_N$ is given formally by \req{2.12}, with respect to $\Delta$.
Next, by inserting the result into \req{3.1} and \req{3.3} one obtains the density
of condensed triplons $\rho_0$ and the glassy fraction $\rho_G$, respectively. The
total density can be evaluated then by the normalization condition $\rho=\rho_0+\rho_N+\rho_G$.
Equations \req{3.7a}, \req{3.7b} for the superfluid density are formally the same in both approximations.

\section{The shift of the critical temperature due to disorder  and the $T>T_c $ regime}

It is well known that the critical temperature of BEC, $T_{c}$
for an ideal gas is given by:
\be
T_{c}^{0}=\dsfrac{2\pi}{m}\;\left(\dsfrac{\rho_{c} }{\zeta(3/2)}\right)^{2/3},
\lab{4.1}
\ee
where $\rho_{c}$ is the total density of triplons near the critical temperature of
BEC for pure system,
\be
\rho_{c}=\mu/2U,
\lab{4.2}
\ee
 with $\zeta(x)$ being the Riemann function. Equation \req{4.2} directly follows, from
 Eqs. \req{2.6}, \req{2.7}  or  \req{3.2} by setting  $\rho_N=\rho$ and $\rho_0=\rho_G=0$.

Clearly, any type of interaction is expected to modify $T_c$. In general, these modifications
are related to the interparticle interactions as well as to the disorder potential.
Both approaches, considered here give a zero shift due to the boson-boson repulsion.
However the shift due to the $\delta-$correlated disorder \req{2.18},
$\Delta T_c=T_c-T_{c}^{0}$ is given as \ci{yukgraham,yukechaya}
\be
\dsfrac{\Delta T_c}{T_{c}^{0}}=-\dsfrac{2\nu}{9\pi},
\lab{4.3}
\ee
where the  dimensionless disorder parameter $\nu $
\be
\nu\equiv \dsfrac{1}{\rho_{c}^{1/3} L_{\rm loc}},
\lab{4.2nu}
\ee
 is introduced with the localization length
 \be
 L_{\rm loc}=\dsfrac{4\pi}{7m^2R}.
 \lab{4.2loc}
 \ee
For practical calculations we rewrite  $T_c$ in Eq. \req{4.3}, which is in a good agreement with perturbative
estimates \ci{vinokur} as well as with Monte Carlo simulations
\ci{pilati},  as an explicit function of effective mass $m$,
the interaction strength  $U$, critical magnetic field
$H_c$, disorder parameter $\nu$, and  external field $H$ as follows:
\be
T_c=\dsfrac{9\pi-2\nu}{9m}\left(\dsfrac{\sqrt{2}g\mu_B (H-H_c)}{U\zeta(3/2)}\right)^{2/3}.
\lab{4.4}
\ee
Now we proceed to consider the triplon density in the normal state in the $T-T_c\gg\Delta T_{c}$ temperature
range. The dirty bosons in the normal phase where the gauge symmetry is not broken,
are yet poorly studied. For $R=0$ with $\rho_0=\rho_G=\sigma=0$ the triplon gas
behaves like  an "ideal gas" with an effective chemical potential $\mu_{\rm eff}$, and the density
\ci{nikinu} 
  \be
\ba
\rho(T>T_{c})=\dsint \dsfrac{d^3{k}}{\left( 2\pi \right) ^{3}}
\dsfrac{1}{\exp \left((\veps_{k}-\mu_{\rm eff})/T\right) -1}.
\lab{5.1}
\ea
\ee
Although $\mu_{\rm eff} $ is not  accurately
   known it  depends  in general, on $\rho$, as well as on $R$. For the pure
  case MFA \ci{nikinu} gives $\mu_{\rm eff}(R=0)=\mu-2U\rho$.
   The contribution from the  disorder potential
  has  been studied neither in YM nor in HM approaches.
    Therefore, to make the calculations self consistently, we have to use
   \be
\ba
\rho(T>T_{c})=\dsint \dsfrac{d^3{k}}{\left( 2\pi \right) ^{3}}
\dsfrac{1}{\exp \left((\veps_{k}-\mu-2U\rho )/T\right) -1},
\lab{5.2}
\ea
\ee
which yields the density $\rho$ as a solution of the nonlinear equation \req{5.2}.

\section{Results and discussions}

In the calculations below, the energies are measured in Kelvin, the mass in K$^{-1}$,
the densities are dimensionless and the Bohr magneton is $\mu_B=0.671668$ K/T.
As to the strength of disorder potential $R$,
it has units K$^{-2}$ while the disorder parameter $\nu$, defined in Eq. \req{4.2nu} is a number supposed to be less
than one, $\nu<1$. As a material parameter, we use mean dimer-dimer distance in TlCuCl$_3$ ${r}_{dd} =0.79$ nm \ci{Amore08}.

To perform numerical calculations in the YG approximation, assuming that $\mu$, $U$, $m$, and $R$ are given parameters, we use following strategy.
(i) By inserting  \re{2.6}, \re{2.11}, \re{2.19} and  \re{2.20} into \re{2.10} we  obtain quadratic algebraic equation
with respect to $\rho$ and solve it analytically. (ii) By using this $\rho(\mu,R,\Delta)$ and \re{2.12}, \re{2.13} in \re{2.16}
we solve the latter  numerically  with respect to $\Delta$, and (iii) by inserting this
$\Delta$ back  into $\rho(\mu,R,\Delta)$ we find the magnetization from \re{2.9} and evaluate other densities
like $\rho_0$ and $\rho_G$ from \re{2.19} and \re{2.20}.

\begin{figure}[h]
\includegraphics*[width=0.65\textwidth]{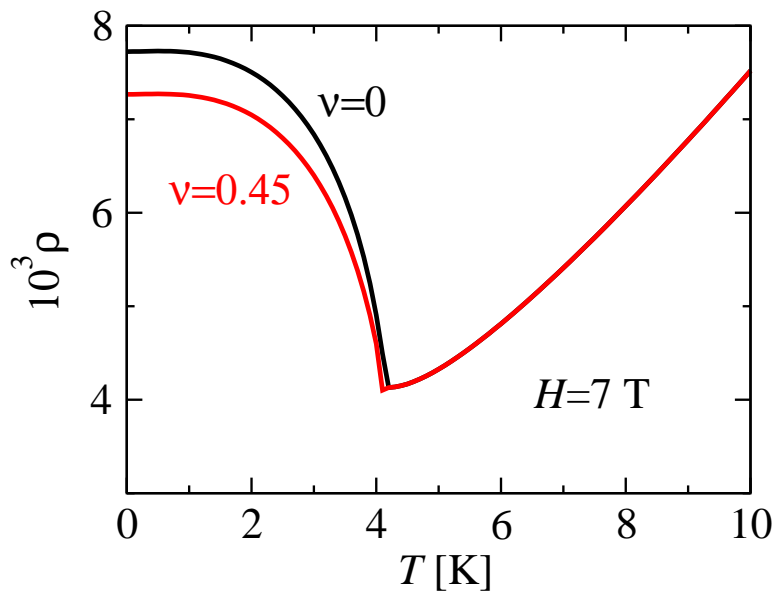}
\caption{The total triplon density
as a function of temperature in the YG approximation for two values of $\nu$. Here the following set of parameters
$m$=0.0204 $\mbox{K}^{-1}$,  $\Delta_{\rm st}=7.3$ K,  $U=313$ K, and $g=2.06$ \ci{Yamada08}  valid for TlCuCl$_{3}$  is used.}
\label{fig1}
\end{figure}

 In Figure \ref{fig1} we present  as an example the total triplon density
 $\rho(T)$ for a clean and strongly disordered ($\nu=0.45$, see Eq.\re{4.2nu}) TlCuCl$_3$, obtained in
 the YG approximation assuming that the total effect of the doping leads only to the randomness
in the triplon subsystem.

\begin{figure}[h]
\includegraphics*[width=0.65\textwidth]{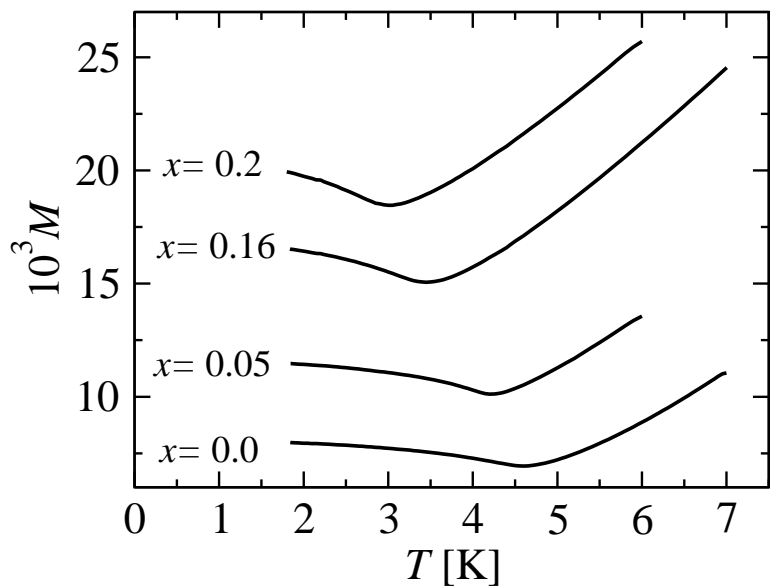}
\caption{
The experimental low temperature magnetization in units of Bohr magneton per Cu ion of Tl$_{1-x}$K$_x$CuCl$_3$ for various $x$ in $H=7$ T
magnetic field obtained in Ref.\ci{oosawa65}.
}
\label{fig2}
\end{figure}

The calculation of other quantities using the same assumption shows that the disorder leads to a decrease
in the condensed and superfluid fractions, thereby increasing the glassy one.
This tendency is quite natural, since the localization effects prevent particles from
going into BEC. However, the increase in $\rho_G$ is so weak that along with
$\rho_0$ the total number of triplons $\rho$ is also decreased with increasing the strength of
disorder potential $R$. Bearing in mind that $\rho$ is proportional to the magnetization $M$,
and $\nu$ is assumed to be approximately proportional to $x$, and comparing Fig.\ref{fig1} with the experimental
data illustrated in Fig.\ref{fig2} one may conclude that the agreement
between the theory and the experiment is unsatisfactory since the main features of the experimental results
are not reproduced there. As it is seen in Fig.\ref{fig2} the disorder leads to an increase in the magnetization and,
hence, in the total triplon density. This is accompanied by the
decrease in the transition temperature. We therefore conclude that while the triplon gas can be considered
similarly to atomic gases for which the  considered mean-field approximations were developed, some
further additional specific material - related  properties of the  dirty boson
problem in quantum magnets must be taken into account.

First we note that the singlet - triplet excitation gap $\Delta_{\rm st}$, proportional
to  the critical field, $H_c$, decreases under high pressure. This was
experimentally observed in Ref.\ci{tanakapressure} for the pure spin system TlCuCl$_3$.
On the other hand it can be argued that
the doping acts as a chemical pressure, which decreases $H_c$.
In fact, since the ionic radius of K$^{+}$ is smaller than that of Tl$^+$,
a partial  substitution of Tl$^+$ ions with  K$^+$
ions produces not only the exchange randomness, but also a compression of
the crystal lattice. Thus the increase of the doping parameter, $x$,  leads to decrease in $H_c$ which
has indeed been observed experimentally \ci{oosawa65,shindotanaka,filho}.
Second, the disorder may increase the triplon effective mass thereby
decreasing the critical temperature $T_c$ even when the gap decreases
(similar effects were observed for helium in porous media \ci{chan88,shibayama}).
Note that this effect manifests itself  in different  ways.
For example, for the mixed compound IPACu(Cl$_x$Br$_{1-x}$)$_3$ the critical field,
$H_c$ remains almost unchanged with varying $x$ and then, abruptly becomes zero near the Cl-rich phase
\ci{manaka2002}. In another triplon-BEC compound, Ni(Cl$_{1-x}$Br$_{x}$)$_2$-4SC(NH$_2$)$_2$,
it decreases by a factor of two when $x$ changes from zero to 0.08 \cite{filho} although the physics
of this decrease can be different from that in Tl$_{1-x}$K$_{x}$CuCl$_3$ due to the fact that
Br atomic radius is larger than the atomic radius of Cl.
 These effects of renormalization of the triplon spectrum by disorder
can be considered similarly to the virtual crystal approximation in the simulations of disorder
in solids, where the disorder is assumed to lead to a uniform change in the system parameters. The effects
of disorder such as the appearance of the glassy phase with the density $\rho_G$ and related phenomena 
manifest themselves in addition to
these uniform changes.

The phase diagram of Tl$_{1-x}$K$_x$CuCl$_3$ in the $(H,T)$ plane  was
experimentally determined in Refs.\ci{oosawa65,shindotanaka} for various doping  $x$,
and the critical field $H_c$  was also estimated by extrapolation to zero temperature.
In  the present work the  $T_c (H)$ dependence
is given by Eq.  \req{4.4}. We  made an attempt  to least - square fit our parameters $m$ and $\nu$
by using Eq. \req{4.4} to describe the experimental phase diagram.
For simplicity we assume that interparticle interaction is not changed by doping, i.e.  $U=U(R=0)=313$ K.
The parameters obtained by this optimization are presented in Table 1.

\begin{table}[ tp]%
\caption {Optimized values of the input parameters of the model:
the critical field, $H_c$ (taken from Ref. \ci{oosawa65}),
 the disorder parameter $\nu$,  and the effective mass $m$ for various
doping $x$. The critical density, $\rho_c$, the healing length,
$\lambda=1/\sqrt{2m\mu}$,  the interparticle distance, $d=1/\rho_{c}^{1/3}$
and the localization length, $L_{\rm loc}=d/\nu$ are estimated at $H=7$ T. It is assumed that
the doping effects does not modify the Land\'e factor $g$ and $U$.
 }
\begin{tabular}{|c|c|c|c|c|c|c|c|c|}
  \hline
  $x$&
    $\nu$  &
     $ H_c$ [T] &
      $m$ [1/K] &
       $ \Delta_{\rm st}$ [K]&
        $\rho_{c}$ &
         $\lambda $ [nm] &
          $ d $ [nm] &
           $L_{\rm loc}$ [nm] \\
  \hline
   0   &0     &5.3  & 0.020  & 7.3& 0.0038 & 2.55 & 5.08 &$\infty $   \\
  \hline
  0.05 &0.16 & 4.8 &0.024  & 6.6& 0.0049 & 2.04 & 4.64 & 28.4 \\
  \hline
  0.08 & 0.25& 4.4 & 0.029  & 6.1&0.0059  & 1.71 & 4.38 & 17.7  \\
  \hline
  0.16 & 0.48& 4.1 & 0.039  & 5.6&0.0065  & 1.40 & 4.23 & 8.86  \\
  \hline
  0.20  & 0.59& 3.9 & 0.044  & 5.4&0.0068  & 1.28 & 4.16 & 7.09  \\
  \hline
\end{tabular}
\end{table}

\begin{figure}[h]
\includegraphics*[width=0.9\textwidth]{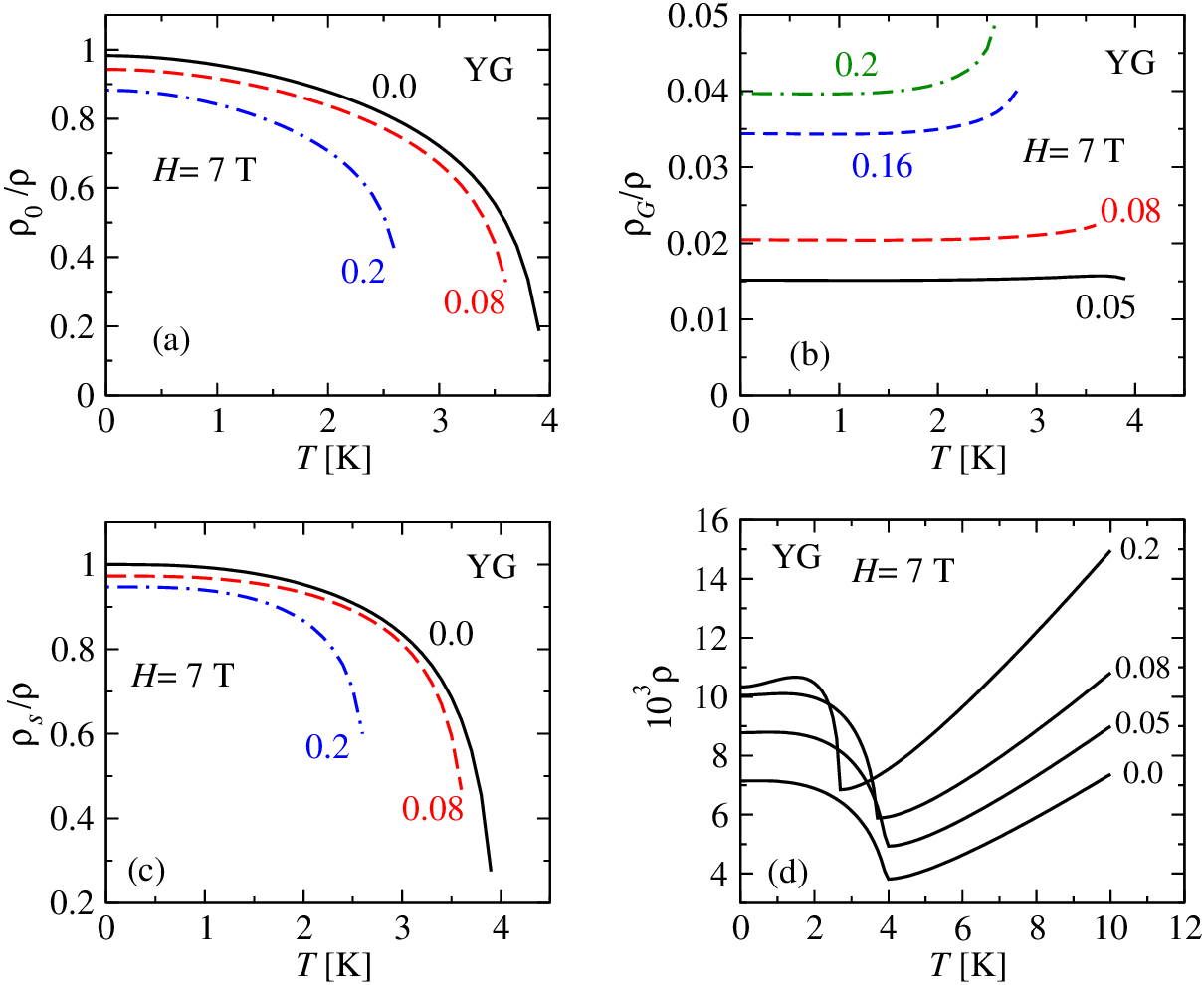}
\caption{The condensed (a), glassy (b), and superfluid fractions (c) as functions of temperature in
the YG approximation for various $x$ marked near the plots with the input parameters from Table 1.
The total density of triplons is shown in Fig.3(d).}
\label{fig3}
\end{figure}

Having fixed the input parameters for certain values of $x$, we are now in the position
of recalculating the densities as well the magnetization to compare them  with the experiment.
Figure \ref{fig3} shows that the doping decreases $\rho_0$ and $\rho_s$, and  increases
$\rho_G$ as it is expected due to the introduced disorder.
Due to change of $H_c$ with $x$, the total density of triplons and hence
the magnetization, now  increases with increasing  $x$ in accordance with the experiment.
One may conclude that the YG approach may well  describe  the effect of disorder to the magnetization,
with the additional assumption of an  $x$ dependence of the effective mass and the critical field.

\begin{figure}[h]
\includegraphics*[width=0.45\textwidth]{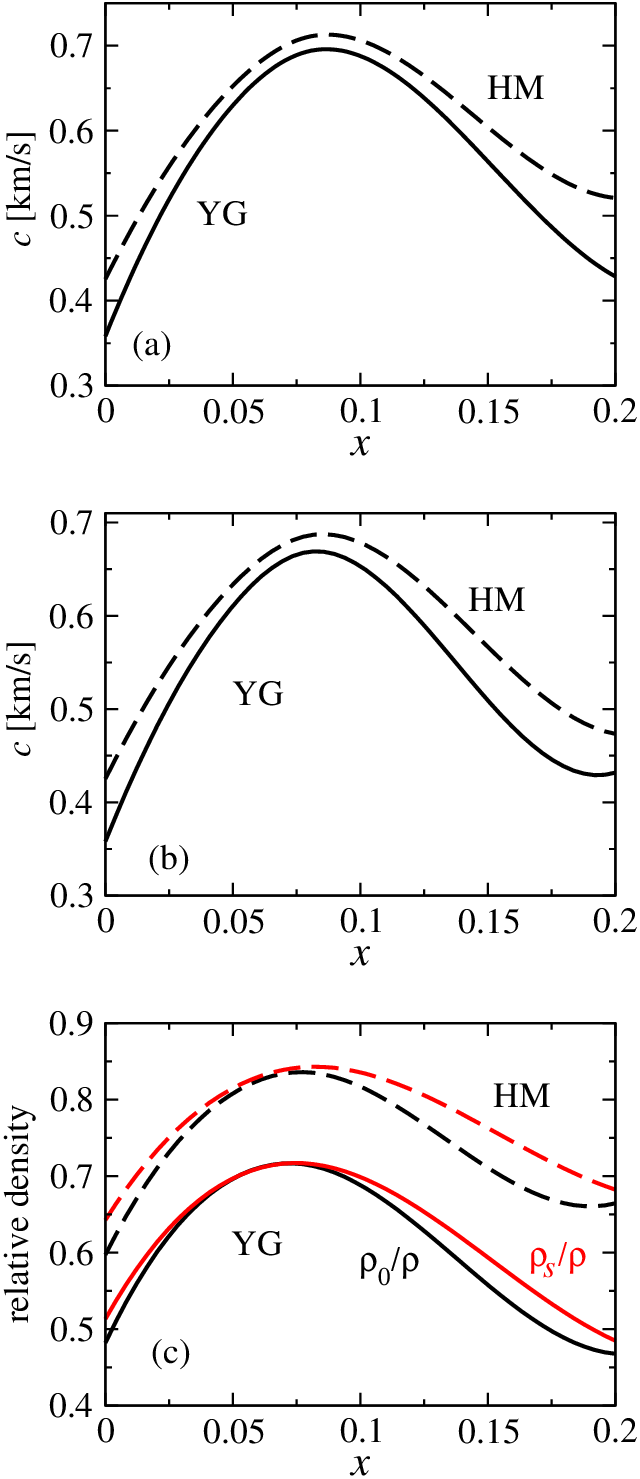}
\caption{{(a) The speed of sound-like condensate mode at $T=2$ K and  $H=6$ T as a function of doping parameter $x$,
taking into account solely renormalization of the system parameters in Table 1 in the YG (solid line) and the
HM (dashed line) approaches. (b) The same as in Fig. 4(a), now with effects of disorder taken into account.}
(c) The superfluid and condensed fractions (as marked near the plots) in the YG (solid lines) and the HM (dashed lines)
approximations with the renormalized bare spectrum parameters.}
\label{fig4}
\end{figure}

{One of the main characteristics of Bose condensed  systems is the speed of the Bogoliubov mode $c$,
defined here by the Eq. \req{2.15}, which characterizes the propagation of collective excitations in the condensate.
It is interesting to mention that the magnitude of $c$ is large, being only an order of magnitude less
than the speed of sound in the crystal. This is due to very small triplon mass in TlCuCl$_{3}$.} Clearly,
disorder modifies the small-momentum excitation spectrum of the BEC. Estimates
of such modification, $\Delta c=c-c_0$,  where $c_0$ is the speed of the Bogoliubov mode for the pure system,
 that exist in the literature are controversial.
For example, perturbative \ci{pelster} and hydrodynamic \ci{giorgini} approaches
 give $\Delta c>0$,  while $\Delta c<0$ was predicted in Refs. \ci{gaul,zhang}.
In Fig. \ref{fig4} we present the corresponding speed for various doping parameters.
{It can be seen from comparison of Fig.\ref{fig4}(a) and Fig.\ref{fig4}(b) that both MFA approximations considered here
show the decrease in $c$ with increasing the disorder strength due to the localization effects. However, the effect
of disorder is small leading to a less than 10 percent decrease in the speed of the Bogoliubov mode.}

However, when the spectrum modification by disorder is also taken into account
 by a renormalization of the triplon mass and the gap, as close to the real situation,
 the dispersion of the sound-like mode in fixed magnetic field slightly increases
with increasing disorder, reaches a maximum and then  starts to decrease, (see Fig.\ref{fig4}b).
This behavior is caused by interplay  between
renormalization of the system parameters and localization effects.
The former tends to increase $c$, e.g. by increasing $\mu$, and therefore increasing the density,
while the latter tends to decrease $c$, e.g. by decreasing the condensed fraction.
Note that an increase in $c$ with increasing  the  density was experimentally observed by Andrews {\it et al.} \ci{andrews}
for the BEC of sodium atoms.
This interplay is illustrated  in
 Fig.\ref{fig4}(c) for $\rho_0$ and $\rho_s$. It can be  seen that uniform spectrum renormalization first
  leads to an antidepletion effect, increasing these quantities,
  while the localization effects impair the  condensation and superfluidity.

We now consider the question about the existence of a pure Bose glass phase at $T=0$,
which, strictly speaking, should fulfill the following criteria \ci{yukgraham,fisher,Yukalovobsor}:
(i) gapless in the excitation spectrum, (ii) insulating behavior, i.e. the superfluid fraction, $\rho_s=0$,
(iii) finite compressibility, and (iv) finite density of states.

\begin{figure}[h]
\includegraphics*[width=0.9\textwidth]{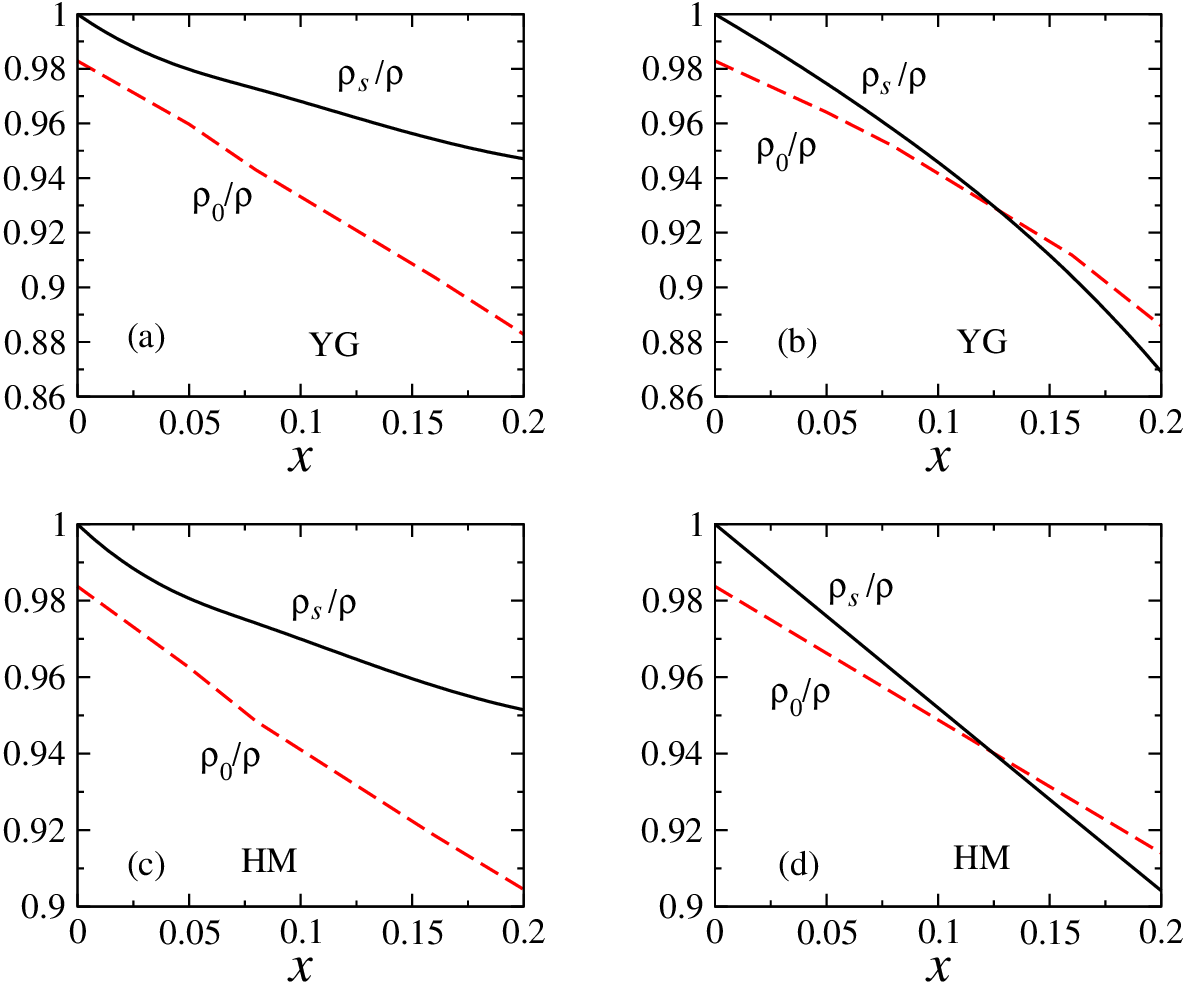}
\caption{The  superfluid, $\rho_s/\rho$ (solid lines) and  condensed, $\rho_0/\rho$ (dashed lines),
fractions as a function of the doping parameter $x$  at $T=0$, $H=7$ T.
Upper panel corresponds to the YG approximation and lower panel corresponds to the HM approximation
presented here for comparison.
Graphs in plots (a), (c) were calculated without bare spectrum renormalization, while graphs in plots (b), (d)
take into account spectrum renormalization presented in Table 1.}
\label{fig5}
\end{figure}

In 1970, Tachiki and Yamada \ci{tachiki} have shown that the Heisenberg-like Hamiltonian of $s=1/2$
dimers can be rewritten as an effective bosonic Hamiltonian.
Recently, Roscilde and Haas \ci{roscilde} generalized this bosonization procedure taking into account
disorder and derived a Bose - Hubbard like Hamiltonian usually applied
to study "dirty bosons" in optical lattices.  Applying
Fishers ideas \ci{fisher} we may expect the formation of a pure Bose glass phase for doped magnets
such as Tl$_{1-x}$K$_x$CuCl$_3$. Although  Monte Carlo calculations \ci{nohadani,roscilde}
confirmed its existence the experimental confirmation is still a matter of debate
\ci{discusplr,hongzheludev}. We underline    here  that these Bose glass phases are  localized
out of the BEC phase, i.e. for $H<H_c$. However, in the  present work we have
been mainly concentrating on the region with $H\geq H_c$
where the gapless phase can be realized only within the  BEC phase. For this case the definition
of BG phase may be simplified as a phase with $\rho_0\neq 0$ and $\rho_s=0$ since
the spectrum of the  BEC is gapless by itself.  In
searching for such a phase we studied $\rho_s$ and $\rho_0$ at $T=0$ for various $H\geq H_c$
and $x$ and found no pure BG phase with $\rho_s=0$ as illustrated in Fig. \ref{fig5}.

Note also that, as it  is seen from Table 1, for moderate values of $x$ considered here,  
the localization length, i.e   the mean free path \ci{falcoprb80}
is larger than interparticle distance, $L_{\rm loc}>d$.
The physics of the possible BG phase for $H<H_c$ fields will be the subject of a separate study.

\section{Conclusions}

We reformulated and applied two existing mean field approximations
for the  "dirty boson" problem to study properties of Tl$_{1-x}$K$_x$CuCl$_3$ quantum magnets.
We showed that these approaches can qualitatively explain the magnetization
data if a certain  modification of the model parameters 
similar to the virtual crystal approximation usually applied to electron spectrum of alloys is taken into account.

 In fact, random bond effects in mixed dimerized magnetic compounds
   manifest themselves in a dual way: (i) by modification of internal parameters and
   (ii) by  localization on random  scatterers.
   Each of these effects could be studied  separately in an appropriate
   theory, but they should be taken into account simultaneously for an adequate
   description of the measured magnetization data. Although the system
becomes considerably disordered, the Bose-Einstein condensation does not support formation of the pure Bose glass phase.
      The random bonds lead to a nontrivial behavior of the sound-like mode speed:
   when $H$  is  fixed and $x$ is experimentally varied, it increases
   for small $x$,  reaches a maximum value and then  decreases.
    While the speed of this mode was measured in dilute BEC
   of sodium atoms a long time ago \ci{andrews}, it has never been an intense focus of research
    in dimerized quantum magnets \ci{andreas}. It could be systematically studied, for example, by measuring the dispersion
    relation of the Bogoliubov mode with inelastic neutron scattering techniques.

\section*{Acknowledgments}
The  work is partly supported
by the Schweizerische Nationalfonds zur F\"orderung
der wissenschaftlichen Forschung under Grants No. IZK0Z2-139441 and No.20-140465.
The work of EYS was supported by the MCINN of Spain grant FIS
2009-12773-C02-01, "Grupos Consolidados UPV/EHU del Gobierno Vasco" grant
IT-472-10, and by the UPV/EHU under program UFI 11/55.
We are indebted to H. Tanaka for providing us with tabulated data for Fig.\ref{fig2}.


\bb{99}
\bibitem{pelster} Falco G M,  Pelster A and Graham R  2007  {\it Phys. Rev.} A {\bf 75} 063619 
\bibitem{vinokur} Lopatin A V and  Vinokur V M  2002  {\it Phys. Rev. Lett.}  {\bf 88} 235503 
 \bibitem{giorgini}  Giorgini S,  Pitaevskii L and  Stringari S  1994  {\it Phys. Rev.} B {\bf 49} 12938 
 \bibitem{gaul}  Gaul C,  Renner N and M\"{u}ller C A   2009  {\it Phys. Rev.} A {\bf 80} 053620;
  Gaul C and   M\"{u}ller C A  2011  {\it Phys. Rev.} A {\bf 83} 063629 
 \bibitem{zhang}  Zhang L  1993  {\it Phys. Rev.} B {\bf 47} 14364.
\bibitem{graham} Graham R and Pelster A 2009 {\it Int. Journ. of Bifurcation and Chaos} {\bf 19} 2745
\bibitem{shelykh}  Shelykh I A,  Kavokin A V,  Rubo Y G,  Liew T C H and Malpuech G  2010  {\it Semicond. Sci. Technol.} {\bf 25} 013001
and references therein.
\bibitem{butov}   Butov L V,   Gossard A C and Chemla D S  2002  {\it Nature}  {\bf 418} 751 
\bibitem{shapiro} See recent topical review:  Shapiro B   2012  J. Phys. A {\bf 45} 143001
\bibitem{yukgraham}  Yukalov V I and  Graham R {\it Phys. Rev.} A {\bf 75} 023619  2007 ;
 Yukalov V I,  Yukalova E P,  Krutitsky K V and  Graham R  2007  {\it Phys. Rev.} A {\bf 76} 053623 
\bibitem{huangmeng} K. Huang and H. F. Meng  1992  {\it Phys. Rev. Lett.}  {\bf 69} 644 
\bibitem{demokritov}  Demokritov S O,  Demidov V E,  Dzyapko O,  Melkov G A,  Serga A A,  Hillebrands B and  Slavin A N
                 2006  {\it Nature}  {\bf 443} 430 
\bibitem{giamarchi}     Giamarchi T,   R\"{u}egg C and Tchernyshyov O  2008  {\it Nature  Phys.} {\bf 4} 198 
\bibitem{oosawa65}  Oosawa A and  Tanaka H   2002  {\it Phys. Rev.} B {\bf 65} 184437 
\bibitem{yamadaglass}  Yamada F,  Tanaka H,  Ono T and  Nojiri H  2011  {\it Phys. Rev.} B {\bf 83} 020409 
\bibitem{tanakaptp}  Tanaka H,  Shindo Y and  Oosawa A   2005  {\it Progress of Theoretical Physics Supplement} 159, 189 
\bibitem{Yamada08}  Yamada F,  Ono T,  Tanaka H,  Misguich G,  Oshikawa M and  Sakakibara T 
              2008  {\it J. Phys. Soc. Jpn.}  {\bf 77} 013701 
\bibitem{ourmagnon}  Rakhimov A,  Mardonov S and  Sherman  E Ya  2011  {\it Ann. Phys.}  {\bf 326} 2499;
  Rakhimov A,  Sherman E Ya and Kim C K  2010  {\it Phys. Rev.} B {\bf 81} 020407  
\bibitem{Amore08}  Dell'Amore R,  Schilling A, and  Kr\"{a}mer K  2009 
             {\it Phys. Rev.} B {\bf 79} 014438; Dell'Amore R,  Schilling A and  Kr\"{a}mer K
              2008  {\it Phys. Rev.} B {\bf 78} 224403 
\bibitem{discusplr} Zheludev  A and  H\"{u}vonen D  2011  {\it Phys. Rev.} B {\bf 83} 216401;
  Yamada F,  Tanaka H,  Ono T and  Nojiri H 2011  {\it Phys. Rev.} B {\bf 83} 216402
 \bibitem{monte}  Astrakharchik G E,  Boronat J,  Casulleras J and Giorgini S  2002  {\it Phys. Rev.} A {\bf 66} 023603 
 \bibitem{nohadani}  Nohadani O,  Wessel S and  Haas S  2005  {\it Phys. Rev. Lett.}  {\bf 95} 227201 
 \bibitem{yukannals} Yukalov V I  2008  {\it {\it Ann. Phys.} } {\bf 323} 461 
 \bibitem{fisher}  Fisher M P A,  Weichman P B,  Grinstein G and Fisher D S  1989  {\it Phys. Rev.} B {\bf 40} 546 
\bibitem{yukechaya}  Yukalov V I  2011  {\it Physics of Particles and Nuclei} {\bf 42} 460 
\bibitem{pilati} Pilati S,  Giorgini S and Prokofev N  2009  {\it Phys. Rev. Lett.}  {\bf 102} 150402 
\bibitem{nikinu}  Nikuni T,  Oshikawa M,  Oosawa A and  Tanaka H  2000  {\it Phys. Rev. Lett.}  {\bf 84} 5868 
\bibitem{tanakapressure} Tanaka H,  Goto K,  Fujisawa M,  Oho T and  Uwatoko Y  2003  {\it Physica} B {\bf 329-333} 697;
Goto  K,  Fujisawa M,  Tanaka H,  Uwatoko Y,  Oosawa A,  Osakabe T and  Kakurai K
 2006  {\it J. Phys. Soc. Jpn.}  {\bf 75} 064703;   Goto K,  Osakabe T,  Kakurai K,  Uwatoko Y,  Oosawa A,  Kawakami J and  Tanaka H
 2007  {\it J. Phys. Soc. Jpn.}  {\bf 76} 053704 
\bibitem{shindotanaka} Shindo Y and  Tanaka H  2004   {\it J. Phys. Soc. Jpn.}  {\bf 73} 2642 
\bibitem{filho}  Yu R {\it et al.} 2012 Nature \textbf{489} 379;  Paduan-Filho A  2012 arXiv:1206.0035
 \bibitem{chan88}  Chan M H W,  Blum K I,  Murphy S Q, Wong G K S and  Reppy J D
  1988  {\it Phys. Rev. Lett.}  {\bf 61} 1950 
\bibitem{shibayama}  Shibayama Y and  Shirahama K  2011  {\it J. Phys. Soc. Jpn.}  {\bf 80} 084604 
 \bibitem{manaka2002}  Manaka H, Yamada I,  Mitamura H and  Goto T  2002  {\it Phys. Rev.} B {\bf 66} 064402 
\bibitem{andrews}  Andrews M R,  Kurn D M,  Miesner H-J,  Durfee D S,  Townsend C G,  Inouye S and  Ketterle W
            1997 {\it Phys. Rev. Lett.}  {\bf 79} 553
 \bibitem{Yukalovobsor}  Yukalov V I  2009  {\it Laser Physics}  {\bf 19} 1 
 \bibitem{tachiki}  Tachiki M and  Yamada T  1970  {\it J. Phys. Soc. Jpn.}  {\bf 28} 1413 
 \bibitem{roscilde} Roscilde T  and  Haas S.  2006  J. Phys. B {\bf 39} S 153 
 \bibitem{hongzheludev} Hong T, Zheludev A, Manaka H and Regnault L-P  2010  {\it Phys. Rev.} B {\bf 81} 060410  
 \bibitem{falcoprb80}  Falco G M,  Nattermann T and  Pokrovsky V L   2009  {\it Phys. Rev.} B {\bf 80} 104515 
 \bibitem{andreas} Schilling A, Grundman H and Dell'Amore R 2011 arXiv:1107.4335
 \eb

\edc